\documentclass[lettersize,journal]{IEEEtran}
\usepackage{amsmath,amsfonts}
\usepackage{textgreek}
\usepackage{algorithmic}
\usepackage{algorithm}
\usepackage{array}
\usepackage[caption=false,font=normalsize,labelfont=sf,textfont=sf]{subfig}
\usepackage{textcomp}
\usepackage{xcolor}
\usepackage{stfloats}
\usepackage{url}
\usepackage{verbatim}
\usepackage{graphicx}
\usepackage{cite}
\usepackage{authblk}
\usepackage{graphicx}

\DeclareMathOperator*{\argmax}{arg\,max}

\usepackage{ifthen}
\usepackage{amssymb}
\usepackage{xcolor}

\newboolean{showcomments}
\setboolean{showcomments}{true}

\makeatletter
\newcommand{\mynote}[3]{%
  \ifthenelse{\boolean{showcomments}}{%
   \fbox{\bfseries\sffamily\scriptsize#1}%
   {\small$\blacktriangleright$\textsf{\emph{\color{#3}{#2}}}$\blacktriangleleft$}}%
  {%
   \@bsphack
   \@esphack
  }%
}
\makeatother

\newcommand{\ze}[1]{\mynote{Zephan}{#1}{black}}

\setboolean{showcomments}{false}

\begin{document}

\title{A 65nm Privacy-preserving Neuromorphic Encoder with 7.13\,nJ Efficiency, 2.38\,Mb/mm\textsuperscript{2} 2T-2T Item Memory Density and Federated Learning Support}

\author{Boyang Cheng,~\IEEEmembership{Student Member,~IEEE,} Jianbo Liu,~\IEEEmembership{Student Member,~IEEE,}
    Steven Davis,~\IEEEmembership{Student Member,~IEEE,} Zephan M. Enciso,~\IEEEmembership{Student Member,~IEEE,}
    Likai Pei,~\IEEEmembership{Student Member,~IEEE,} Xueji Zhao,~\IEEEmembership{Student Member,~IEEE,} Muya Chang,~\IEEEmembership{Member,~IEEE,} 
    Ningyuan Cao,~\IEEEmembership{Member,~IEEE}
\thanks{This work has been submitted to the IEEE for possible publication. Copyright may be transferred without notice, after which this version may no longer be accessible.}
    }
\markboth{}%
{Shell \MakeLowercase{\textit{et al.}}: A Sample Article Using IEEEtran.cls for IEEE Journals}

\maketitle

\begin{abstract}

The increasing demand for privacy-preserving personal data analytics in smart assistants and context-aware systems necessitates hardware that is both energy-efficient and secure. This work presents a 65\,nm privacy-preserving neuromorphic encoder that addresses these challenges through physically uncloneable and hardware-efficient encoding. At the core of the architecture is a 2T-2T entropy cell that enables compact and uncloneable hyperdimensional representations. The fabricated prototype achieves 7.13\,nJ per encoding, 2.38\,Mb/mm\textsuperscript{2} item memory density, and supports in-situ decision-making and continual learning with 76.44\,nJ per prediction and 357.32\,nJ per training update. To support multi-user deployment, a custom federated learning framework is proposed, enabling efficient cold-start personalization while preserving device-level model privacy. Evaluations on bio-signal datasets demonstrate strong classification accuracy—93.2\% on EMG and 96.1\% on UCI-HAR—with additional gains of up to xx$\times$ through federated training.

\end{abstract}

\begin{IEEEkeywords}
Federated learning, hyper-dimensional computing, bio-signal processing.
\end{IEEEkeywords}

\section{Introduction}

\begin{figure}[tp]
    \centering
    \includegraphics[width=0.9\linewidth]{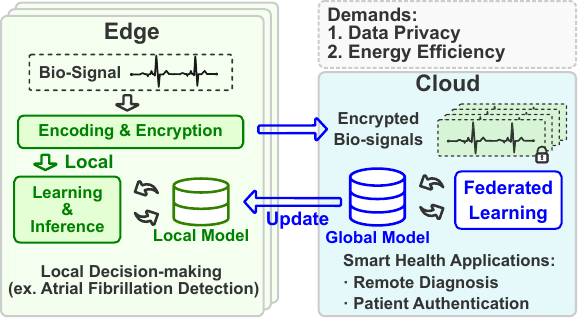}
    \caption{Overview of federated learning enabled edge-cloud collaboration for bio-signal applications.}
    \label{fig:motivation}
    \vspace{-0.4cm}
\end{figure}

\IEEEPARstart{P}{rivacy-preserving} personal data analytics is becoming increasingly critical in emerging applications such as smart digital assistants, wearable health monitors, and biometric authentication systems. These systems process sensitive user data—including electromyography (EMG), electrocardiography (ECG), and fingerprint patterns—requiring hardware that ensures both energy efficiency and data confidentiality. For instance, Sole et al.~\cite{Sole_EMG} proposed an EMG pre-processing accelerator achieving up to 1.2\,MS/s for multichannel signal digitization. Agarwala et al.~\cite{Agarwala_ECG} developed a 785\,nW ECG interface for real-time cardiovascular monitoring, while Chang et al.~\cite{chang_fingerprint} introduced a fingerprint authentication SoC with 792.5\,ms anti-spoofing latency. These examples underscore the growing need for personalized yet secure on-device analytics. However, conventional software-based security solutions—such as homomorphic encryption~\cite{yang_fpga-based_2020} and differential privacy~\cite{hassan_differential_2020}—incur substantial hardware overhead, making them impractical for resource-constrained platforms that demand immediate and energy-efficient decision-making (Fig.~\ref{fig:motivation}).

Hyperdimensional computing (HDC) is a neuromorphic computing model based on operations on high-dimensional vectors \cite{kanerva_hyperdimensional_2009}. Its distributed vector representations enable effective learning from limited data, making it particularly suitable for biomedical applications where data availability is often constrained. Additionally, HDC’s parallel and relatively simple computations make it compatible with compute-in-memory (CIM) acceleration hardware. Several HDC accelerators using CIM across various memory types (e.g., SRAM \cite{imani_low-power_2017}, ReRAM \cite{Karunaratne2020}, FeRAM \cite{kazemi_achieving_2022}, and PCM \cite{karunaratne_real-time_2021}) have demonstrated improvements in hardware efficiency.

Despite these advantages, HDC faces notable challenges. Generating and storing hyper-dimensional random vectors in item memory (IM) for data encoding incurs significant hardware overhead, limiting its scalability \cite{Liu2025_IoTJ}. Cheng et al. \cite{Cheng_VAE_HDC} proposed a 5T-5T macro based HDC encoder which greatly reduces energy consumption by direct encoding in analog domain. However, it still faces a relatively high area overhead. Moreover, the straightforward encoding operations, such as simple multiplication and addition, make the model vulnerable to privacy risks, as raw data can potentially be reconstructed by accessing the IM \cite{Liu_hyperattack}. These limitations pose significant barriers to adopting HDC in privacy-critical and resource-constrained applications.

To address these challenges, we present a neuromorphic bio-signal encoder chip that leverages transistor-based physically unclonable entropy. The main contributions are listed as the following:

\begin{enumerate} 


\item \textbf{Physically-Unclonable Encoding}: Ensured data privacy through unique, physically-unclonable encoding and device-specific HDC item memory.

\item \textbf{Federated Learning Framework}: Integrated federated learning with physically-unclonable HDC to enable secure, collaborative learning across edge devices. This framework improves the accuracy for clients with limited data, facilitates a warm start for new nodes, and shows a typical classification accuracy improvement of 14.8\%. 

\item \textbf{Accuracy and Scalability}: Demonstrated competitive algorithmic performance with 93.2\% accuracy on EMG and 96.1\% on UCIHAR datasets. Federated learning further enhances performance and scalability across diverse bio-signal datasets.

\item \textbf{Low Memory Footprint} By leveraging compact fabrication randomness in HDC encoding, it achieves energy-efficient bio-signal encoding at 7.13 nJ and a item memory density of 2.38 Mb/mm\textsuperscript{2}.

\end{enumerate}

\section{Hyperdimensional Computing}
\label{sec:HDC}

\begin{figure*}[htp]
    \centering
    \includegraphics[width=0.9\linewidth]{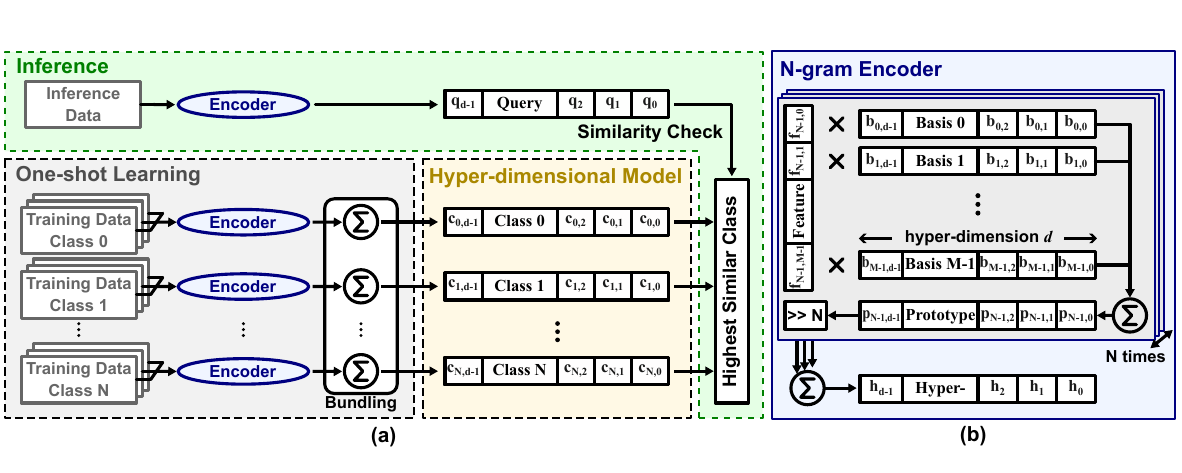}
    \caption{Hyperdimensional computing overview. (a) The low-dimensional training data is encoded to form class Hyper-vectors (one-shot training), and then perform similarity check with the query vector encoded by same encoder, the class with the highest similarity is the result of inference. (b) HDC n-gram encoder: N pieces of input data are vector-matrix multiplied with the basis vectors in item memory and then accumulated to obtain an encoded hyper-vector.}
    \label{fig:hyper-dimensional_computing}
    \vspace{-0.4cm}
\end{figure*}

In this section, we introduce and give an overview on hyperdimensional computing (HDC). Next, we discuss the learning and encoding mechanism of HDC.

\subsection{Background}
\ze{Gently altered for flow and length.} HDC is an emerging paradigm inspired by brain-like properties such as high dimensionality, distributed representation, and pseudo-randomness. It has shown promise in tasks like speech recognition, anomaly detection, and bio-signal processing due to its low data requirements \cite{imani_framework_2019}. In certain machine learning (ML) tasks, HDC achieves comparable or superior accuracy to traditional models (e.g., SVMs, CNNs) while consuming less energy \cite{imani_framework_2019}.
\ze{Likewise altered} In HDC, the $d$--dimensional hyper-vectors are (pseudo)--random and contain independent and identically distributed (\textbf{i.i.d}) elements.  When $d$ is large (typically greater than 1000), the hyper-vectors are quasi-orthogonal, so they can be combined into new, unique hyper-vectors using well-defined vector space operations \cite{karunaratne_real-time_2021}. \ze{Why is this important?}

\subsection{N-gram HDC Encoding}
\label{sec:ngram}
The first step of HDC in training and inference is \emph{encoding} (see Fig.~\ref{fig:hyper-dimensional_computing}). Encoding maps the low-dimensional input feature vector (e.g. time-series stream, image) onto a hyper-vector. Given an input feature vector $\mathbf{f} = [f_0, f_1, \dots, f_{M-1}]^T,\ \mathbf{f} \in \mathcal{F}$, encoding $\mathbf{f}$ maps it to its high-dimensional representation $\mathbf{h} \in \mathcal{H}$ with dimension $d \gg M$ using transformation $map: \mathcal{F} \rightarrow \mathcal{H}$. There are several mapping algorithms such as random projection \cite{Papadimitriou1998}, permutation \cite{kanerva_hyperdimensional_2009}, and position-ID \cite{Thomas2021}; this work employs N-gram permutation because it better protects the spatial information compared to the other approaches \cite{kleyko_holographic_2017}.

To perform N-gram permutation, $N$ adjacent feature vectors $[\mathbf{f_{i,0}}, \mathbf{f_{i,1}}, \dots, \mathbf{f_{i, M-1}}],\ i = 0, 1,\dots, N-1$ are multiplied with a matrix $\mathbf{B} \subset \mathbb{R}^{N \times d}$, which consists $M$ basis vectors $\mathbf{b}_i$. Since all elements in basis vectors are randomly generated \textbf{i.i.d.} from a certain distribution, these basis vectors are quasi-orthogonal; that is, $\mathbf{b}_i \cdot \mathbf{b}_j \approx 0, (i \neq j) $. After the mapping, each feature vector is transformed to its corresponding prototype vector $\mathbf{p}_i$ \ze{how?} and ready for N-gram binding. 
To prevent the spatial information between feature vectors, the prototypes perform a permutation which is an element-wise shifting. \ze{What does this mean?} Finally, all shifted prototypes vectors are multiplied together to form a single hyper-dimension:


\begin{equation}
    \mathbf{h} = \prod^{N-1}_{i=0} \rho^i \cdot \mathbf{p}_i, ~ i = [0, 1, \dots, N-1]
\end{equation}

\begin{equation}
    \mathbf{p}_i = \mathbf{f}_i \cdot \mathbf{B}, ~ i = [0, 1, \dots, N-1]
\end{equation}

\ where $\rho$ is circular shift operator and $\rho^i$ represents a $i$-step circular shifting, and $\prod$ means a element-wise vector multiplication.
With this \textbf{i.i.d.}-preserving binding, features are combined to \textbf{i.i.d.} hypervectors for both inference and one/few-shot learning, as shown in Fig.~\ref{fig:hyper-dimensional_computing}~(b).

\subsection{HDC One-shot Learning and Inference}
\label{sec:oneshot}
\ze{Signficantly shortened} Although many machine learning tasks perform well with HDC, we focus on one-shot supervised learning, as shown in Fig.~\ref{fig:hyper-dimensional_computing}(a). Given a set of labeled data $\mathcal{D} = \{ (\mathbf{x}_i,y_i) \}^n_{i=0}$, where $\mathbf{x}_i \in \mathcal{X} \subset \mathbb{R}^M$ and $y_i \in \mathcal{Y}$ is an enumerated variable indicating the class of the corresponding sample, the training examples $\mathbf{x_i}$ are first encoded into hyper-vectors $\mathbf{h_i}$ as in Sec. \ref{sec:ngram}.  Then, all encoded examples with the same $k$--th class are bundled together into class vector $\mathbf{c_k}$:

\begin{equation}
    \mathbf{c}_k = \sum_{i~|~y_i=y_k} \mathbf{h}_i
\end{equation}

Bundling all $N+1$ classes in the training dataset yields the HDC model $\mathbf{C}=[\mathbf{c}_0, \mathbf{c}_1, \dots, \mathbf{c}_N]$.  Since the entire dataset is only accessed once without iterations, this training method can be regarded as one-shot learning.
\ze{Does this jive with the definition of one shot? I just want to double check}

Inference of HDC has a two-step procedure. The first step encodes the input (similar to encoding during training) to produce a query hyper-vector $\mathbf{q} = map (\mathbf{f}_q)$.
Then, the query hyper-vector is compared to each of the class hyper-vectors.  This work uses cosine similarity, a commonly used metric in HDC, which measures the angle between two hyper-vectors and returns a real value between $0$ (dissimilar) and $1$ (similar):

\begin{equation}
    \hat{y_q} = \argmax_{k=[0, \dots, N]} \frac{\langle \mathbf{c}_k , \mathbf{q} \rangle}{||\mathbf{c}_k||}
\end{equation}

The class with the highest similarity score is selected as the predicted label. Cosine similarity is computationally efficient, making it suitable for hardware implementations of HDC. By reducing complex operations into simple vector dot products and norm calculations, HDC systems can achieve fast, accurate inference while maintaining energy efficiency.

\begin{figure}[tp]
    \centering
    \includegraphics[width=0.9\linewidth]{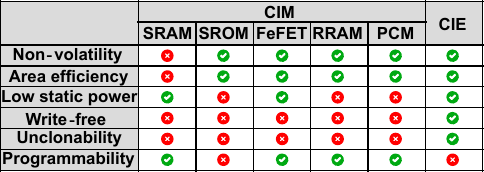}
    \caption{Compute-in-Entropy approach compared with state-of-the-art approaches.}
    \label{fig:CIE_comparison_table}
    \vspace{-0.4cm}
\end{figure}

\subsection{HDC Challenges}

Given the high dimensionality of HDC item memory (typically $>$1000), conventional SRAM-based implementations suffer from substantial energy and area overhead for generating and storing hypervectors. Additionally, SRAM’s volatility limits its applicability in always-on and energy-constrained systems. To address these limitations, alternative non-volatile memory technologies such as FeFET~\cite{li2024}, RRAM~\cite{wu2018}, and PCM~\cite{karunaratne_real-time_2021} have been explored, offering better energy efficiency and smaller area footprints. However, these solutions still incur considerable energy costs when generating and writing basis hypervectors, making them less suitable for low-power applications.

Beyond efficiency concerns, the linear and deterministic nature of HDC introduces significant privacy risks. If basis hypervecctors (in item memory) and query hypervectors are exposed, raw feature values can be reconstructed using $|x_m| = \frac{\mathbf{Q} \cdot \mathbf{B}_m}{N}$, where $N$ is the hyperdimension~\cite{davis2024}. This makes the system vulnerable to various physical and algorithmic attacks, including side-channel~\cite{Liu_side_channel_attacks}, fault injection~\cite{Barenghi_fault_injection_attacks}, and bit-flipping attacks~\cite{Liu_bit_flipping_attacks}. Furthermore, unlike deep neural networks that use non-linear operations to obscure raw data representations, HDC encoding is more transparent and may be reverse-engineered to reveal sensitive information, undermining data privacy.

To address these challenges, we adopt the compute-in-entropy (CIE) architecture proposed in \cite{pei2025}, which integrates transistor-based physically unclonable entropy as the HDC item memory within a CIM framework. This approach enables energy- and area-efficient, device-specific, privacy-preserving bio-signal encoding. Building on this structure, we demonstrate federated learning using physically unclonable encoding, allowing secure training initialization for new patients or users while ensuring that all data remains protected by a device-specific root of trust.

\section{Variation-based HDC}

In this section, we discuss the vulnerability of HDC, and demonstrate a privacy-preserving variation-based compute-in-entropy hyper-dimensional computing (CIE-HDC).

\subsection{Compute-in-Entropy Hyper-dimensional Computing}
To address the limitations of high write energy and security vulnerabilities, we propose a compute-in-entropy (CIE) approach. Inspired by physical unclonable functions (PUFs), CIE leverages inherent transistor-level variations from fabrication as the entropy source for HDC encoding. Compared to memory-based storage of basis hypervectors, CIE offers significant advantages in efficiency, security, robustness, and hardware simplicity—at the cost of reduced programmability (Fig.~\ref{fig:CIE_comparison_table}).

To address the privacy concerns, CIE-HDC directly extracts analog entropy from process variation and use them as to generate the basis hyper-vectors matrix $\textbf{B} = [\textbf{b}_0^T, \textbf{b}_1^T, \dots, \textbf{b}_{N-1}^T]$. CIE functions similarly to non-volatile memory. However, no real data is stored on-chip, which mitigates potential information leakage. Furthermore, due to die-to-die process variations, each device exhibits a unique fingerprint, serving as a root of trust in the HDC encoding process. Once the transformation from low-dimensional to hyper-dimensional space is complete, it becomes challenging for attackers to recover the original data from the encoded hyper-vectors without access to the specific fingerprint. Additionally, compromising one device does not reduce the privacy of others, since the physical unclonability.

\subsection{Algorithm Evaluation}

\begin{figure}[t]
    \centering
    \includegraphics[width=0.8\linewidth]{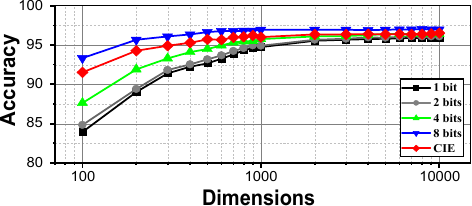}
    \caption{Accuracy on EMG dataset for different hyper-dimensions. 1, 2, 4, 8-bit represent the element inside the hyper-vectors' precision of digital baseline.}
    \label{fig:accuracy_vs_dimensions}
    \vspace{-0.4cm}
\end{figure}

Each element in CIE-HDC structure is represented by a continuous quantity, so the equivalent bit precision for each element is higher than binary HDC. Therefore, the dimensions can be significantly reduced compared to the conventional binary HDC. To evaluate the dimensions saving, we test CIE-HDC using EMG dataset comparing with conventional digital approaches. Fig.~\ref{fig:accuracy_vs_dimensions} shows the accuracy on EMG dataset for different hyper-dimensions. We can see that to achieve the same accuracy CIE approach has a significant dimension saving. For example, to achieve 96\% accuracy, CIE saves 14.3$\times$ dimensions compared with binary (1-bit) approach. Meanwhile, since most of the computation happen in analog domain, CIE framework only uses 4 transistors to store an element inside hyper-vectors. Compared with a conventional SRAM based compute-in-memory system, our approach greatly reduces the required devices resource by more than 6$\times$.

\subsection{Attack Resistance \& Data Reversibility}

\begin{figure}[t]
    \centering
    \includegraphics[width=0.95\linewidth]{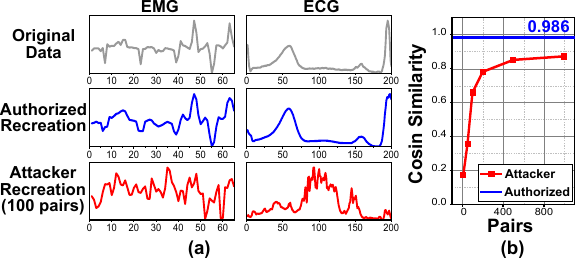}
    \caption{(a) Demonstration of data reversibility. (b) Data reconstruction quality comparison.}
    \label{fig:attack_resistance}
    \vspace{-0.3cm}
\end{figure}

For certain complex applications requiring additional computational resources or expert analysis from a trusted third party, maintaining the reversibility of the original data is desirable. In such cases, the hyper-vector can be shared along with its fingerprint, enabling authorized users to reconstruct the original data. Fig.~\ref{fig:attack_resistance}~(a) presents results for both authorized users and attackers. Here, we simulate a known input-output pair attack, assuming that malicious users attempt to utilize leaked information to infer the fingerprint. We can see that the authorized users' reconstructions maintain the key features of the signals. However, with 100 input-output pairs leaking, the attackers still can not reconstruct the original data. As shown by Fig.~\ref{fig:attack_resistance}~(b), with more pairs known by the attackers, the quality of the reconstructions increases but will never reach the authorized reconstructions, which means it can protect the sensitive information from being used by malicious three parties.

\section{Federated Hyper-dimensional Computing}

\label{sec:federated_learning}

Incorporating data from multiple clients into the training pipeline creates more accurate and robust models, and federated learning with HDC enables a learning framework that aggregates client data without compromising privacy.

\begin{figure}[tp]
    \centering
    \includegraphics[width=0.8\linewidth]{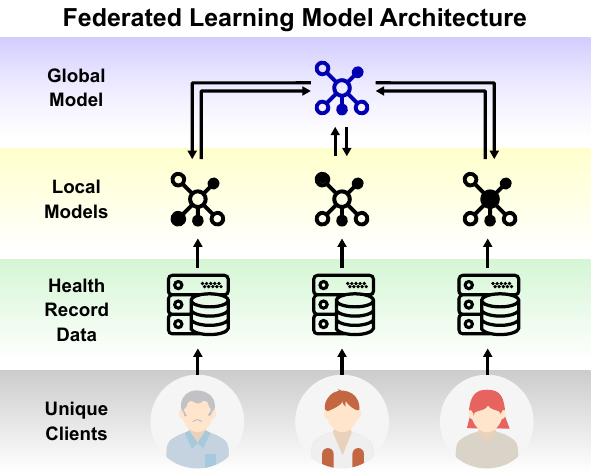}
    \caption{Federated learning architecture.}
    \label{fig:federated_learning_arch}
    \vspace{-0.3cm}
\end{figure}

\subsection{Federated Learning}
\ze{shortened} Federated learning (FL) is a computing paradigm that allows privacy-preserving, collaborative learning across edge devices \cite{Zhao_federated_learning}.  As shown in Fig.~\ref{fig:federated_learning_arch}, clients perform local model training, and a central server aggregates the shared models before broadcasting the new global model back to existing and new clients.  In this way, participating clients all contribute to improving the global model, but each client's collected data remains local.  In this work, FL assists edge nodes with insufficient local data and facilitates a warm start for newly added nodes in the network. \ze{if they start with data, the start is no longer cold}

\subsection{Federated HDC}
\begin{figure}[tp]
    \centering
    \includegraphics[width=1\linewidth]{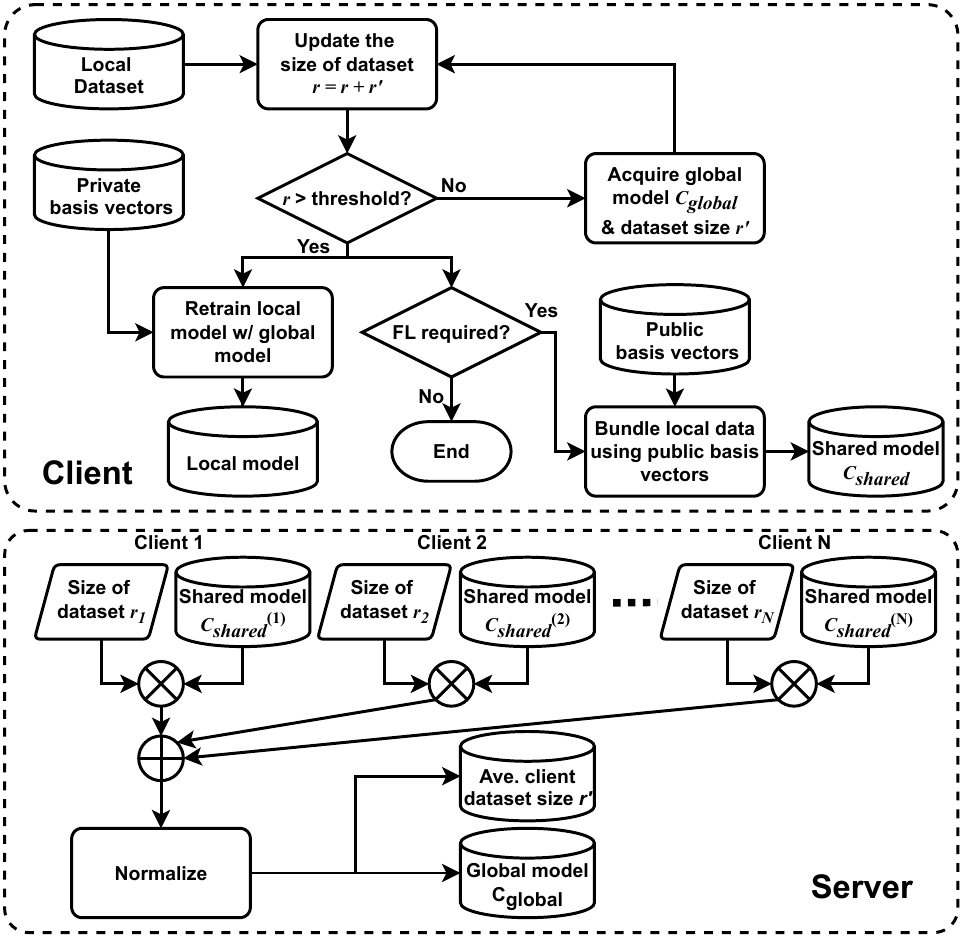}
    \caption{Federated learning flowchart.}
    \label{fig:FL_flowchart}
\end{figure}

The flowchart of client local HDC is shown by Fig.~\ref{fig:FL_flowchart}. After a collection of bio-signal samples is gathered in the local dataset on the edge, the client first updates the size of the local dataset $r$. If $r$ exceeds a pre-set threshold, indicating that the local dataset contains sufficient training data to achieve satisfactory accuracy, the client can perform regular HDC as described in Sec.~\ref{sec:HDC}. If $r$ falls below this threshold, the client will request an FL global model. Upon receiving this request, the server broadcasts it, prompting the relevant clients to bundle a portion of their data using public basis vectors to generate shared models $\textbf{C}_{shared}^i$. Since different client have a private HD basis hyper-vectors based on process variation, direct data sharing cannot provide direct help to the global model. To facilitate federated learning, each client needs to store a public HD basis vectors to learn the shared model.

Similarly, each joint client initially starts the process with a local on-chip training by bundling all encoded hyper-vectors with the same class together using the public basis hyper-vectors which generates corresponding class vectors $\mathbf{c}_k=\sum^{n}_{i} \rho^i \cdot \mathbf{h}^k_i$. The the shared model is a set of hyper-vectors with the number of classes in the local dataset. We define the shared HD model $\mathbf{C}_{shared}=[\mathbf{c}_s^1,\mathbf{c}_s^2,\dots,\mathbf{c}_s^\alpha]$. This model $C_{shared}$ will be shared with the trusted third party along with the amount of local data $r$.

After a trusted third party collects data from clients, federated learning is performed as shown in Eq.~\ref{eq:FL}.
\begin{equation}
\label{eq:FL}
    C_{global} = \sum^N_{i} r_i \cdot \textbf{C}_{shared}^i
\end{equation}
where $r_i$ is size of each client's shared dataset, $N$ indicates the number of participating client. As shown by Fig.~\ref{fig:FL_flowchart}, after FL, the global model is transmitted back to each client along with the average size of shared dataset $r'$. 
The client which receives the global model, it updates the size of local dataset $r=r+r'$
If the average data amount exceeds the threshold, the client accepts the global model and converts it to local model with a matrix $\textbf{B}^*$. This $\textbf{B}^*$ is a pseudo-inverse matrix of a local private HD basis hyper-vectors and can be calculated by linear regression. Client transforms it into a local model using a local transformation matrix $\textbf{B}^{*}$ as shown by Eq.~\ref{eq:pseudo_inverse} 
\begin{equation}
    \label{eq:pseudo_inverse}
    \textbf{c}_l^k = (\textbf{c}_g^k \cdot \textbf{B}^*)^T
\end{equation} 
where $\textbf{c}_l^k$ indicates the k-th class in global model, $\textbf{c}_g^k$ indicates the k-th class in local model. Since this process occurs only when local data is insufficient or during a cold start, it is typically performed only once per edge node, making its impact on the node's battery life negligible.

\begin{figure}[tp]
    \centering
    \includegraphics[width=0.8\linewidth]{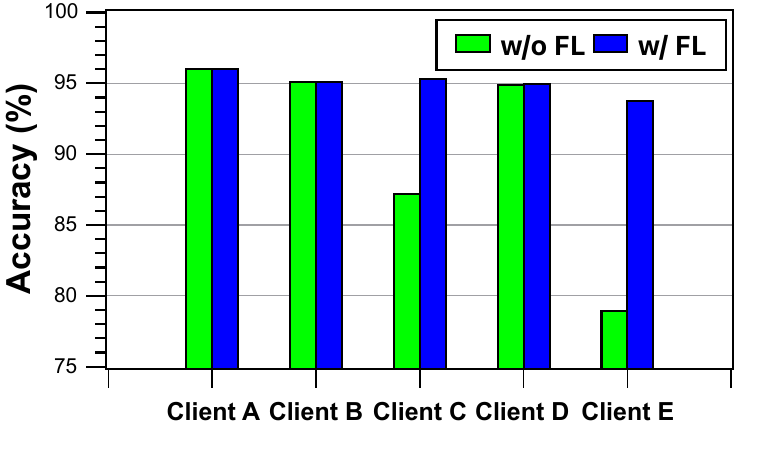}
    \vspace{-0.5cm}
    \caption{Accuracy comparison across clients with different local dataset size, Client A: 1100 samples, Client : 1100 samples, Client C: 200 samples, Client D: 1100 samples, Client E: 50 samples (threshold = 1000).}
    \label{fig:FL_accuracy}
\end{figure}

Fig.~\ref{fig:FL_accuracy} presents a comparison of accuracy between scenarios with and without federated learning (FL) when the data is insufficient. There are five clients in a FL collaboration. Clients A and B have 50 and 200 samples, respectively. Since clients C, D, and E have local datasets exceeding the threshold, they continue using their local models. We observe that federated learning significantly enhances the accuracy of clients with limited samples, bringing it close to the accuracy achieved by clients with sufficient data.

\section{Architecture and Circuits Design}
\label{sec:architecture_and_circuits}

Typically, process variation introduces unwanted variability or degrades circuit performance.  By taking advantage of the randomness offered by process variation, Compute-in-Entropy enables efficient HDC encoding, learning, and inference.


\subsection{Compute-in-Entropy Cell}

\begin{figure}[t]
    \centering
    \includegraphics[width=0.9\linewidth]{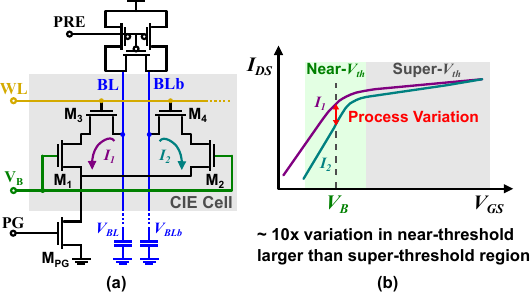}
    \caption{(a) Compute-in-Entropy cell with pre-charger and power-gating footer, (b) drain-source current difference due to process variation in near-threshold region is approximately 10 times larger than that is super-threshold region.}
    \label{fig:CIE_cell}
\end{figure}

The CIE cell (Fig.~\ref{fig:CIE_cell}~(a)) consists of a two-transistor differential pair M1/M3 and M2/M4.  In each operating cycle, the two paired bitlines BL and BLb are first precharged to $V_{DD}$. Then, enabling the wordline (WL) causes gating transistors M3 and M4 to conduct.  The discharging currents $I_1$ and $I_2$ are restricted by M1 and M2, respectively, which are biased in the subtreshold region. Due to process variation (e.g., lithography, doping, and etching), the magnitude of $I_1$ and $I_2$ differ, resulting in a measurable voltage difference across the bitline capacitance.  The differential structure cancels any spatial correlation between discharge transistors.  Each row of CIE cells also share a power gating transistor $M_{PG}$, which exploits the CIE cell's nonvolatility to shut off the circuit during the stand-by cycle and further improve energy efficiency.

Biasing the CIE cell to the near-threshold region magnifies process variation by approximately 10 times compared to the super-threshold region (see Fig.~\ref{fig:CIE_cell}~(b)) because the subthreshold drain-source current has an exponential dependence on threshold voltage ($V_{th0}$): \ze{Citation needed?} \ze{Removed ``difficult to measure on mature nodes'', moved comparison}

\begin{equation}
    I_{DS} = I_0 \cdot e^{(\frac{V_{GS} - V_{th0}-\eta V_{DS} + \gamma V_{BS}}{nV_{T}})} \cdot (1 - e^{\frac{V_{DS}}{V_T}})
\end{equation}
\begin{equation}
    V_T = \frac{kT}{q}
\end{equation}

\noindent
where $k$ is the Boltzmann constant, $n$ is the sub-threshold swing coefficient, $\eta$ is the drain--induced barrier lowering (DIBL) coefficient, and $\gamma$ is the body effect coefficient.

\subsection{Compute-in-Entropy HDC}

\begin{figure}[t]
    \centering
    \includegraphics[width=0.9\linewidth]{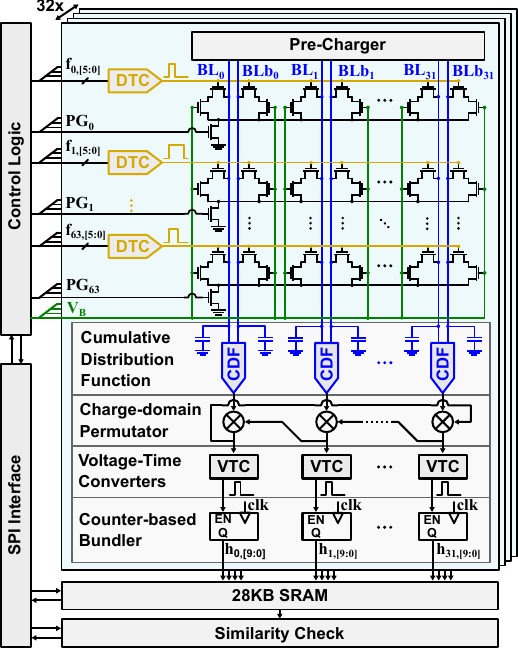}
    \caption{Compute-in-Entropy architecture.}
    \label{fig:CIE_array_arch}
\end{figure}

The CIE core consists of 32 tiles, each with 32$\times$64 CIE cells, which serves as basis vector item memory with a maximum dimension of $d=1024$, as shown in Fig.~\ref{fig:CIE_array_arch}. All tiles share a common feature vector input from the digital control logic circuitry. The maximum feature vector length is 64, and each feature is quantized to 6 bits. A digital-to-time converter (DTC) modulates the digital feature input into a pulse of width $t_i$, which is applied to the WL. The modulated pulse causes the $i$--th CIE cell to conduct sub-threshold current $I_i$, and the resulting voltage difference on the BL capacitance $C_{BL}$ is given by:
\begin{equation}
    \Delta V_{BL} = \sum_{i=0} \frac{I_{i} \cdot t_i}{C_{BL}} , ~ i=[0,1,\dots,63]
\end{equation}


To accommodate the diverse needs of various applications and algorithm, a tunable analog distribution function (CDF) reshapes the original Gaussian--distributed voltage difference to user-specified distributions.
This chip encodes hyper-vectors using N-gram permutation with configurable $N$, executed in the charge-domain permutator. The permutator is followed by a voltage-time converter array, which transforms the hyper-vector, represented by voltage differences from the CIE cells, into time pulses.  Then, a counter-based bundler discretizes the time pulse and performs element-wise accumulation for one-shot learning, as described in Sec. \ref{sec:oneshot}.
The following sections provide detailed descriptions of the aforementioned peripheral circuits.

\subsubsection{Distribution Management}

\begin{figure}[t]
    \centering
    \includegraphics[width=0.9\linewidth]{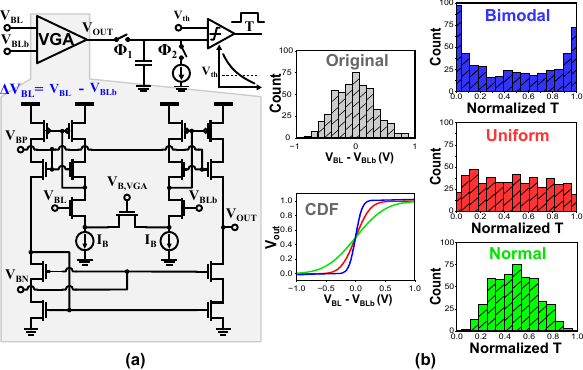}
    \caption{(a) Analog cumulative distribution function. (b) Original normal distribution and different cumulative distribution functions' curves (left). Bimodal, uniform, and normal distribution after distribution reshaping (right).}
    \label{fig:CDF}
\end{figure}

Various applications have different requirements for basis vector distributions, and reshaping distributions typically requires computationally--intensive algorithms that require repeated sampling, such as Monte Carlo Markov Chain \cite{geyer_practical_1992} and acceptance/rejection sampling \cite{accept-reject_sampling}.  Given the stochastic nature of HDC, the transformation does not need to be mathematically exact, so this chip features analog distribution management circuitry to accommodate diverse distribution needs.

The analog commulative distribution function (CDF) consists of two parts, as shown in Fig.~\ref{fig:CDF}~(a):
\textbf{(1)} a variable gain amplifier (VGA) and \textbf{(2)} a charge--based voltage to time pulse converter (VTT). The VGA has differential inputs which connect the BL and BLb. Changing the bias voltage $V_{B,VGA}$ adjusts the transmission curves to mimic different CDFs, mapping the original normal distribution to a desired distribution, as shown by Fig.~\ref{fig:CDF}~(b). The analog CDF also re--normalizes the differential inputs $V_{BL} - V_{BLb}$ to $V_{SS}$ to $V_{DD}$. 
The charge--based VTT then transforms the voltage output from the CDF into a time pulse, beginning when switch $\Phi_1$ opens, allowing the capacitor \ze{label it} to latch the the VGA output voltage $V_{out}$.  Then, switch $\Phi_2$ closes, causing the current mirror to discharge the capacitor until the voltage reaches threshold $V_{th}$, at which point the comparator finishes the output time pulse.

\subsubsection{Charge Domain N-gram Permutation}

\begin{figure}[t]
    \centering
    \includegraphics[width=1\linewidth]{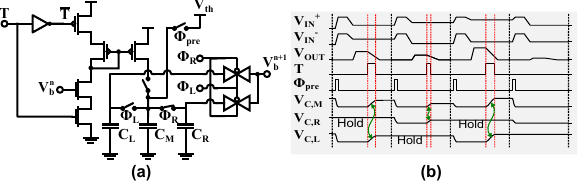}
    \caption{(a) Charge--domain permutator. (b) Original normal distribution and different cumulative distribution functions' curves (left). Bimodal, uniform, and normal distribution after distribution reshaping (right).}
    \label{fig:CDP}
\end{figure}

The charge--domain N--gram permutator seamlessly aggregates $N$ prototypes, as required by the N--gram permutation scheme described in Sec. \ref{sec:HDC}, while eliminating data conversion.
It consists of a voltage-controlled current source (VCCS) followed by an analog buffer, as shown in Fig.~\ref{fig:CDP}~(a). A pair of time pulse signals $T$ and $\overline{T}$ generated by the upstream CDF gate the VCCS.  The output current is controlled by bias voltage $V_B$, which is set by the voltage stored on the analog buffer of the adjacent column.  In N--gram permutation, the prior result is also an input, so the analog buffer contains one primary capacitor $C_M$, charged by the VCCS, and two secondary capacitors $C_L$ and $C_R$, which latch the prior product. The initial voltage $V_b^n$ is set to the NMOS threshold voltage for proper current mirroring. \ze{where?}

Fig.~\ref{fig:CDP}~(b) shows the timing diagram of the charge--domain permutator.
\ze{I think you should label these periods and reference them by name} In the first period, $C_R$ holds the temporary product, while multiplication (between red lines) occurs in $C_M$, and $C_L$ follows the changing of $C_M$. During the next period, $C_L$ latches the temporary product and works as the bias voltage for the adjacent permutator while multiplication occurs with $C_M$ and $C_R$. 
After these cycles, the input hyper-vector is fully encoded and ready for HDC learning and inference.

\section{Post-silicon Measurement Results}

\label{sec:measurement}

\begin{figure}[t]
    \centering
    \includegraphics[width=1\linewidth]{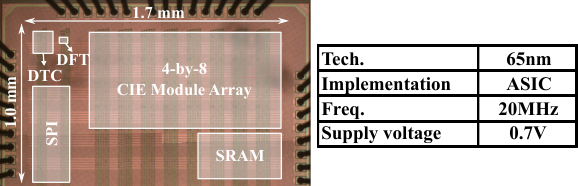}
    \caption{Die micrograph of test chip with 1024 dimensions HDC CIE module array, SRAM array, SPI, digital-to-time pulse converter, and design-for-testing module.}
    \label{fig:die_photo}
\end{figure}

\begin{figure}[t]
    \centering
    \includegraphics[width=0.7\linewidth]{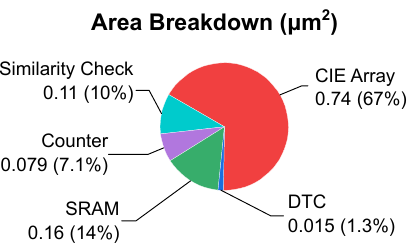}
    \caption{Area breakdown of the test chip.}
    \label{fig:area_breakdown}
\end{figure}

\begin{figure}[t]
    \centering
    \includegraphics[width=0.7\linewidth]{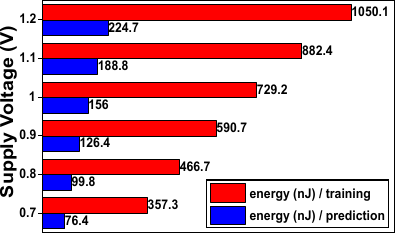}
    \caption{Inference and learning energy across supply voltage.}
    \label{fig:energy}
\end{figure}

A prototype chip on a commercial 65 nm PDK, shown in Fig.~\ref{fig:die_photo}, provides post-silicon training and inference measurements. The test chip includes a 4$\times$8 CIE array, on chip SRAM for buffering hyper-vectors, DTT array, a serial peripheral interface (SPI) for programming and test orchestration, and a design for test module for direct measurements of the CIE cell. The overall chip area is 1.7 mm$^2$, and the CIE arrays occupies over two-thirds of the total area (see Fig.~\ref{fig:die_photo}.  The prototype chip was validated with a 0.7--1.2 V power supply at 20 MHz, as shown in Fig~\ref{fig:energy}. The minimum operating voltage of the prototype chip is 0.7 V. At this voltage, the lowest energy consumption is achieved, which is 357.3 nJ per training and 76.4 nJ per prediction.


\subsection{CIE Cell Evaluation}

\begin{figure}[t]
    \centering
    \includegraphics[width=0.9\linewidth]{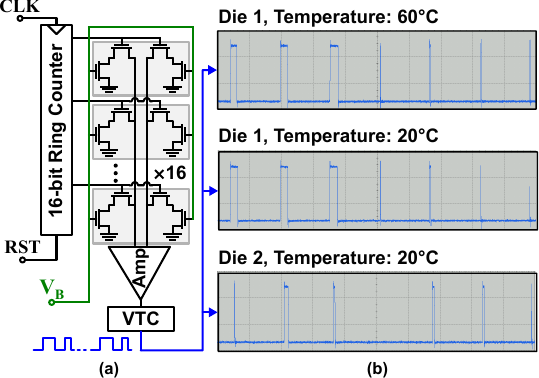}
    \caption{(a) Design-for-test (DFT) module with 16 sequentially activated CIE cells. (b) Examples of generated time pulses from different dies under 20 $^\circ$C and 60 $^\circ$C.}
    \label{fig:design_for_test}
\end{figure}

\begin{figure}[t]
    \centering
    \includegraphics[width=0.9\linewidth]{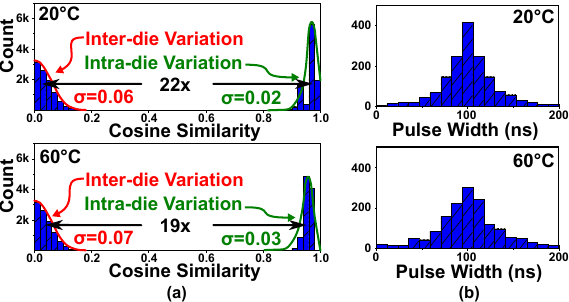}
    \caption{(a) Inter-die cosine similarity shows the uniqueness and intra-die cosine similarity shows the robustness. (b) Time pulses distribution within a typical temperature range of 20 $^\circ$C to 60 $^\circ$C. }
    \label{fig:robustness_and_uniqueness}
\end{figure}

To examine the intrinsic CIE entropy, we build a design-for-test (DFT) module featuring 16 sequentially activated CIE cells on each test chip. Since the process variation is in the form of differential current, we utilize a sense amplifier and voltage to time pulse converter (VTC) to convert the current signals, which are hard to measure, to easily measured digital time pulse, as shown by Fig.~\ref{fig:design_for_test}. After acquiring over 20,000 pulse trains across 100 test chips within a typical temperature range of 20 $^\circ$C to 60 $^\circ$C, which is considered as a working temperature range of most bio-applications, the measurement reveals the robustness, uniqueness, and physical unclonability. Fig.~\ref{fig:robustness_and_uniqueness}~(b) shows the pulse width distributions at 20 $^\circ$C and 60 $^\circ$C. By slightly adjusting the CIE cell bias voltage, both distributions can be tuned to follow a normal distribution with a range of 0 ns to 200 ns and a mean of 100 ns.

For the requirements of robustness, the cosine similarity between response response hyper-vectors for challenges on the same chip should be similar, implying that the intra-die cosine similarity should be close to 1. From Fig.~\ref{fig:robustness_and_uniqueness}~(a) we can see that at 20 $^\circ$C, the intra-die cosine similarity is a half normal distribution with a mean value close to 1 and a 0.02 standard deviation. Meanwhile, at 60 $^\circ$C, the standard deviation increases to 0.03. The result shows that across 20 $^\circ$C to 60 $^\circ$C, CIE has a good robustness.
For the requirements of uniqueness and physical unclonability, the cosine similarity between response hyper-vectors for challenges on different chips should be quasi-orthogonal, which means that the inter-die cosine similarity should be close to 0.
As shown by the red lines in Fig.~\ref{fig:robustness_and_uniqueness}~(a), across the tested temperature range, the inter-die cosine similarity distribution follows a zero-mean half-normal distribution with standard deviations of 0.06 and 0.07.

\begin{table*}[htp]
\centering
\caption{Comparison table across state-of-the-art works}
\begin{tabular}{|l|c|c|c|c|c|c|c|}
\hline
\multicolumn{1}{|c|}{} &
  \textbf{\begin{tabular}[c]{@{}c@{}}TBCAS \\ 2019 \cite{benatti_online_2019}\end{tabular}} &
  \textbf{\begin{tabular}[c]{@{}c@{}}Nature \\  2021 \cite{moin_wearable_2020}\end{tabular}} &
  \textbf{\begin{tabular}[c]{@{}c@{}}TCAS-I\\  2021 \cite{eggimann_5_2021}\end{tabular}} &
  \textbf{\begin{tabular}[c]{@{}c@{}}TCAS-II\\  2021 \cite{karunaratne_real-time_2021}\end{tabular}} &
  \textbf{\begin{tabular}[c]{@{}c@{}}JETCA\\  2019 \cite{datta_programmable_2019}\end{tabular}} &
  \textbf{\begin{tabular}[c]{@{}c@{}}ESSCIRC\\  2023 \cite{datta_hdbinarycore_2023}\end{tabular}} &
  \textbf{This Work} \\ \hline
\textbf{EMG \# of channels} &
  8, diff &
  64, single &
  64, single &
  64, single &
  64, single &
  64, single &
  64, single \\ \hline
\textbf{\# of gestures} &
  11 &
  21 &
  5 &
  5 &
  5 &
  21 &
  5 \\ \hline
\textbf{HDC Dims.} &
  10000 &
  1000 &
  2048 &
  10000 &
  2048 &
  2048 &
  1024 \\ \hline
\textbf{Accuracy (\%)} &
  85 &
  92.87 &
  95.2 &
  98.9 &
  95.5 &
  92.87 &
  \textcolor{red}{\textbf{96.1}} \\ \hline
\textbf{Tech.} &
  40 nm &
  65 nm &
  22 nm &
  90 nm &
  28 nm &
  28 nm &
  65 nm \\ \hline
\textbf{Implementation} &
  MCU &
  FPGA &
  Netlist &
  Netlist &
  Netlist &
  ASIC &
  ASIC \\ \hline
\textbf{Frequency (MHz)} &
  - &
  20.48 &
  1 &
  440 &
  417 &
  200 &
  20 \\ \hline
\textbf{Supply voltage (V)} &
  0.8 &
  1.2 &
  0.6 &
  \begin{tabular}[c]{@{}c@{}}0.1 for PCM\\  1.2 for Netlist\end{tabular} &
  0.8 &
  0.575 &
  0.7 \\ \hline
\textbf{Pred. Throughput (preds./s)} &
  27.8k &
  38.5k &
  147.5 &
  31.5M &
  - &
  44k &
  \textcolor{red}{\textbf{435k}} \\ \hline
\textbf{Memory density (Mb/mm2)} &
  - &
  - &
  - &
  - &
  \begin{tabular}[c]{@{}c@{}}2.68\\  (ROM)\end{tabular} &
  \begin{tabular}[c]{@{}c@{}}0.4\\  (SRAM)\end{tabular} &
  \begin{tabular}[c]{@{}c@{}}\textcolor{red}{2.38}\\  \textcolor{red}{(\textbf{14.04})}\end{tabular} \\ \hline
\textbf{Energy/encoding (nJ)} &
  - &
  - &
  - &
  - &
  - &
  - &
  \textcolor{red}{\textbf{7.13}} \\ \hline
\textbf{Privacy-preserving Encoding} &
  No &
  No &
  No &
  No &
  No &
  No &
  \textcolor{red}{\textbf{Yes}} \\ \hline
\textbf{Memory non-volatility} &
  No &
  No &
  No &
  Yes &
  Yes &
  No &
  \textcolor{red}{\textbf{Yes}} \\ \hline
\textbf{Energy/prediction (nJ)} &
  83,200 &
  1,950 &
  191 &
  13.3 &
  1,500 &
  25.6 &
  \begin{tabular}[c]{@{}c@{}}\textcolor{red}{76.44}\\  \textcolor{red}{(\textbf{21.2nJ})}\end{tabular} \\ \hline
\textbf{Energy/training (nJ)} &
  - &
  - &
  - &
  - &
  - &
  - &
  \begin{tabular}[c]{@{}c@{}}\textcolor{red}{357.32}\\  \textcolor{red}{(\textbf{99.3})}\end{tabular} \\ \hline
\end{tabular}%

\end{table*}
\subsection{Algorithm Evaluation}

Although, CIE framework is mainly focus on the processing of bio-signal, it also has the ability to deal with other kinds of data. To comprehensively evaluate the performance of the proposed CIE framework, various datasets are considered, including bio-signal (EMG \cite{marin_hand_2016} and UCIHAR \cite{jorge_reyes-ortiz_human_2013}), language (ISOLET \cite{ron_cole_isolet_1991} and LANG \cite{rahimi_robust_2016}), and image data (MNIST \cite{lecun_mnist_nodate}). Baselines quantized to 1-, 2-, 4-, and 8-bit precision are implemented for comparison. All implementations have a hyper-dimension of 1024 and the performance of CIE is acquired in room temperature. 

For bio-signal dataset, we take EMG and UCIHAR \cite{jorge_reyes-ortiz_human_2013} as examples and the CIE framework shows a 93.2\% accuracy on EMG dataset and 96.1\% accuracy on UCIHAR dataset. CIE achieves the highest accuracy across all implementations on EMG and between 8-bit and 4-bit baselines on UCIHAR. Language and image datasets evaluations show that CIE has an equivalent bit precession between 4 and 8. Since, it is harder for HDC to encode images than steam data, the accuracy on image dataset is relevantly lower than that on steam data.

\begin{figure}[t]
    \centering
    \includegraphics[width=0.8\linewidth]{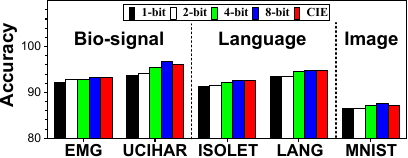}
    \caption{Accuracy evaluation on bio-signal, language, and image datasets. 1, 2, 4, 8-bit represent the element inside the hyper-vectors' precision of digital baseline. All implementations have a dimension of 1024.}
    \label{fig:accuracy}
\end{figure}

\section{Conclusions}
CIE-HDC architecture addresses the privacy and efficiency challenges of smart health applications by combining the benefits of neuromorphic computing and entropy-based encoding. This not only enhances data security, but also supports efficient, real-time bio-signal processing. Compared to traditional HDC approaches, CIE-HDC offers significant advantages, including the generation of physically unclonable basis vectors (BV) through transistor variations, enhancing security against cloning and cyber-attacks. Using analog entropy, it achieves effective data representation with a reduced vector dimension by 14.3$\times$, lowering computational and memory demands. The integration of entropy-based encoding with CIM techniques increases memory density to 2.38 Mb/mm$^2$, making it ideal for wearable bio-signal processors. Leveraging analog entropy enables the application of power gating technology, which further minimizes leakage power—an essential factor for extending the battery life of portable devices. The evaluation of the algorithms shows that the CIE-HDC structure can be well applied to bio-signal algorithms and datasets, with an accuracy of 93.2\% on EMG and 96.1\% on UCIHAR.

\newpage

\bibliographystyle{IEEEtran}
\bibliography{references, zotero}

\vfill

\end{document}